\documentclass[italian,english]{article}
\usepackage[T1]{fontenc}
\usepackage[latin9]{inputenc}
\usepackage{color}
\usepackage{graphicx}
\usepackage{esint}
\usepackage{babel}
\begin{document}

\title{\textbf{The Mössbauer rotor experiment and the general theory
of relativity}}

\author{\textbf{Christian Corda}}

\maketitle
\begin{center}
Dipartimento di Fisica, Scuola Superiore di Studi, Universitari e
Ricerca ``Santa Rita'', Via Tagliamento 45, 00188 Roma, Italy
\par\end{center}

\begin{center}
Austro-Ukrainian Institute for Science and Technology, Institut for
Theoretish Wiedner Hauptstrasse 8-10/136, A-1040, Wien, Austria
\par\end{center}

\begin{center}
International Institute for Applicable Mathematics \& Information
Sciences (IIAMIS),  B.M. Birla Science Centre, Adarsh Nagar, Hyderabad
- 500 463, India 
\par\end{center}

\begin{center}
\textit{E-mail address:} \textcolor{blue}{cordac.galilei@gmail.com} 
\par\end{center}
\begin{abstract}
In the recent paper Eur. Phys. Jour. Plus 130, 191 (2015), the authors
claim that our general relativistic analysis in Ann. Phys. 355, 360
(2015), with the additional effect due to clock synchronization, cannot
explain the extra energy shift in the Mössbauer rotor experiment.
In their opinion, the extra energy shift due to the clock synchronization
is of order $10^{-13}$ and cannot be detected by the detectors of
$\gamma$-quanta which are completely insensitive to such a very low
order of energy shifts. In addition, they claim to have shown that
the extra energy shift can be explained in the framework of the so-called
YARK gravitational theory. They indeed claim that such a theory should
replace the general theory of relativity (GTR) as the correct theory
of gravity.

In this paper we show that the authors Eur. Phys. Jour. Plus 130,
191 (2015) had a misunderstanding of our theoretical analysis in Ann.
Phys. 355, 360 (2015). In fact, in that paper we have shown that electromagnetic
radiation launched by the central source of the apparatus is redshifted
of a quantity $0.\bar{6}\frac{v^{2}}{c^{2}}$ when arriving to the
detector of $\gamma$-quanta. This holds independently by the issue
that the original photons are detected by the resonant absorber which,
in turns, triggers the $\gamma$-quanta which arrive to the final
detector. In other words, the result in Ann. Phys. 355, 360 (2015)
was a purely theoretical result that is completely independent of
the way the experiment is concretely realized. Now, we show that,
with some clarification, the results of Ann. Phys. 355, 360 (2015)
hold also when one considers the various steps of the concrete detection.
In that case, the resonant absorber detects the energy shift and the
separated detector of $\gamma$-quanta merely measures the resulting
intensity. 

In addition, we also show that the YARK gravitational theory is in
macroscopic contrast with geodesic motion and, in turn, with the weak
equivalence principle (WEP). This is in contrast with another claim
of the authors of Eur. Phys. Jour. Plus 130, 191 (2015), i.e. that
the YARK gravitational theory arises from the WEP. Therefore, the
YARK gravitational theory must be ultimately rejected. We also correct
the confusion of the authors of Eur. Phys. Jour. Plus 130, 191 (2015)
concerning their claims about the possibility to localize the gravitational
energy and, in turn, to define a stress-energy tensor for the gravitational
field. In fact, we show that these claims are still in macroscopic
contrast with the WEP.\end{abstract}
\begin{verse}
Paper dedicated to the centenary of the GTR.
\end{verse}

\section{Introduction}

In \cite{key-16} we gave a correct interpretation of a historical
experiment by Kündig on the transverse Doppler shift in a rotating
system measured with the Mössbauer effect (Mössbauer rotor experiment)
\cite{key-3}. The Mössbauer effect (discovered by R. Mössbauer in
1958 \cite{key-14}) consists in resonant and recoil-free emission
and absorption of gamma rays, without loss of energy, by atomic nuclei
bound in a solid. It resulted and currently results very important
for basic research in physics and chemistry. In \cite{key-16} we
focused on the so called Mössbauer rotor experiment. In this particular
experiment, the Mössbauer effect works through an absorber orbited
around a source of resonant radiation (or vice versa). The aim is
to verify the relativistic time dilation for a moving resonant absorber
(the source) inducing a relative energy shift between emission and
absorption lines. 

In a couple of recent papers \cite{key-1,key-2}, the authors first
re-analysed in \cite{key-1} the data of a known experiment of Kündig
on the transverse Doppler shift in a rotating system measured with
the Mössbauer effect \cite{key-3}. In a second stage, they carried
out their own experiment on the time dilation effect in a rotating
system \cite{key-2}. In \cite{key-1} it has been found that the
original experiment by Kündig \cite{key-3} contained errors in the
data processing. A puzzling fact was that, after correction of the
errors of Kündig, the experimental data gave the value \cite{key-1}

\begin{equation}
\frac{\nabla E}{E}\simeq-k\frac{v^{2}}{c^{2}},\label{eq: k}
\end{equation}
where $k=0.596\pm0.006$, instead of the standard relativistic prediction
$k=0.5$ due to time dilatation. The authors of \cite{key-1} stressed
that the deviation of the coefficient $k$ in equation (\ref{eq: k})
from $0.5$ exceeds by almost 20 times the measuring error and that
the revealed deviation cannot be attributed to the influence of rotor
vibrations and other disturbing factors. All these potential disturbing
factors have been indeed excluded by a perfect methodological trick
applied by Kündig \cite{key-3}, i.e. a first-order Doppler modulation
of the energy of $\gamma-$quanta on a rotor at each fixed rotation
frequency. In that way, Kündig's experiment can be considered as the
most precise among other experiments of the same kind {[}4\textendash{}8{]},
where the experimenters measured only the count rate of detected $\gamma-$quanta
as a function of rotation frequency. The authors of \cite{key-1}
have also shown that the experiment {[}8{]}, which contains much more
data than the ones in {[}4\textendash{}7{]}, also confirms the supposition
$k>0.5.$ Motivated by their results in \cite{key-1}, the authors
carried out their own experiment \cite{key-2}. They decided to repeat
neither the scheme of the Kündig experiment \cite{key-3} nor the
schemes of other known experiments on the subject previously mentioned
above {[}4\textendash{}8{]}. In that way, they got independent information
on the value of $k$ in equation (\ref{eq: k}). In particular, they
refrained from the first-order Doppler modulation of the energy of
$\gamma-$quanta, in order to exclude the uncertainties in the realization
of this method \cite{key-2}. They followed the standard scheme {[}4\textendash{}8{]},
where the count rate of detected $\gamma-$quanta $N$ as a function
of the rotation frequency $\nu$ is measured. On the other hand, differently
from the experiments {[}4\textendash{}8{]}, they evaluated the influence
of chaotic vibrations on the measured value of $k$ \cite{key-2}.
Their developed method involved a joint processing of the data collected
for two selected resonant absorbers with the specified difference
of resonant line positions in the Mössbauer spectra \cite{key-2}.
The result obtained in \cite{key-2} is $k=0.68\pm0.03$, confirming
that the coefficient $k$ in equation (\ref{eq: k}) substantially
exceeds 0.5. The scheme of the new Mössbauer rotor experiment is in
Figure 1, while technical details on it can be found in \cite{key-2}. 

\begin{figure}
\includegraphics[scale=0.75]{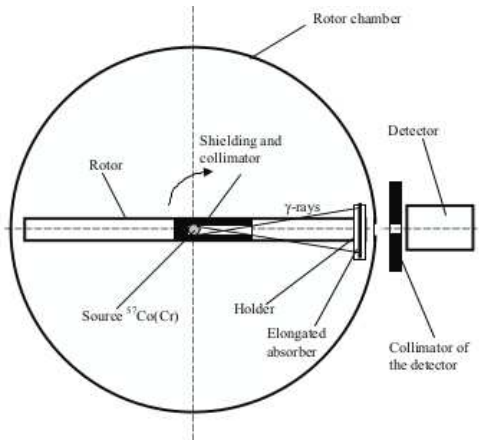}

\caption{Scheme of the new Mössbauer rotor experiment, adapted from ref. \cite{key-2}}
\end{figure}

In \cite{key-16}, the equivalence principle (EP), which states the
equivalence between the gravitational \textquotedbl{}force\textquotedbl{}
and the \emph{pseudo-force} experienced by an observer in a non-inertial
frame of reference (included a rotating frame of reference) has been
used to re-analyse the theoretical framework of Mössbauer rotor experiments.
A full geometric general relativistic treatment was developed directly
in the rotating frame of reference \cite{key-16}. The results have
shown that previous analyses missed an important effect of clock synchronization
and that the correct general relativistic prevision gives $k\simeq\frac{2}{3}$
\cite{key-16}. This was in perfect agreement with the new experimental
results of \cite{key-2}. In that way, the general relativistic interpretation
of \cite{key-16} showed that the new experimental results of the
Mössbauer rotor experiment are a new, strong and independent proof
of the correctness of Einstein's vision of gravity. We also stress
that various papers in the literature (included ref. \cite{key-4}
published in Phys. Rev. Lett.) missed the effect of clock synchronization
\cite{key-1}-\cite{key-8}, \cite{key-11}-\cite{key-13} with some
subsequent claim of invalidity of relativity theory and/or some attempts
to explain the experimental results through ``exotic'' effects \cite{key-1,key-2,key-11,key-12,key-13}. 

In the recent paper \cite{key-17}, it is claimed that the general
relativistic analysis in \cite{key-16}, with the additional effect
due to clock synchronization, cannot explain the extra energy shift
in the Mössbauer rotor experiment. The reason should be that the extra
energy shift due to the clock synchronization is of order $10^{-13}$
and cannot be detected by the detectors of $\gamma$-quanta which
are completely insensitive to such a very low order of energy shifts
\cite{key-17}. In addition, the authors of \cite{key-17} claim to
have shown that the extra energy shift can be explained in the framework
of the so-called YARK gravitational theory. They also claim that such
a new theory should replace the GTR as the correct theory of gravity.

In this paper we show that in \cite{key-17} the authors had a misunderstanding
of our theoretical analysis in \cite{key-16}. In fact, in \cite{key-16}
it has been shown that electromagnetic radiation launched by the central
source of the apparatus is redshifted of a quantity $0.\bar{6}\frac{v^{2}}{c^{2}}$
when arriving to the detector of $\gamma$-quanta. This holds independently
by the issue that the original photons are detected by the resonant
absorber which, in turns, triggers the $\gamma$-quanta which arrive
to the final detector. In other words, the result in \cite{key-16}
was a purely theoretical result that is completely independent of
the way the experiment is concretely realized. Now, we show that,
with some clarification, our results in \cite{key-16} hold also when
we consider the various steps of the concrete detection. In that case
the resonant absorber detects the energy shift and the separated detector
of $\gamma$-quanta merely measures the resulting intensity. 

In addition, in this paper we also show that the YARK gravitational
theory is in macroscopic contrast with geodesic motion and, in turn,
with the WEP. This in contrast with another claim of the authors of
of \cite{key-17}, stating that the YARK gravitational theory should
arise from the WEP. Therefore, the YARK gravitational theory must
be ultimately rejected. We also correct the confusion in \cite{key-17}
concerning the claims about the possibility to localize the gravitational
energy and, in turn, to define a stress-energy tensor for the gravitational
field. In fact, we show that these claims are still in macroscopic
contrast with the WEP.

\section{The ``gravitational'' redshift}

Following \cite{key-9,key-16} let us consider a transformation from
an inertial frame, in which the space-time is Minkowskian, to a rotating
frame of reference. Using cylindrical coordinates, the line element
in the starting inertial frame is \cite{key-9,key-16}

\begin{equation}
ds^{2}=c^{2}dt^{2}-dr^{2}-r^{2}d\phi^{2}-dz^{2}.\label{eq: Minkowskian}
\end{equation}
The transformation to a frame of reference $\left\{ t',r',\phi'z'\right\} $
rotating at the uniform angular rate $\omega$ with respect to the
starting inertial frame is given by \cite{key-9,key-16}

\begin{equation}
\begin{array}{cccc}
t=t'\; & r=r' & \;\phi=\phi'+\omega t'\quad & z=z'\end{array}.\label{eq: trasformazione Langevin}
\end{equation}
Thus, eq. (\ref{eq: Minkowskian}) becomes the following well-known
line element (Langevin metric) in the rotating frame \cite{key-9,key-16}
\begin{equation}
ds^{2}=\left(1-\frac{r'^{2}\omega^{2}}{c^{2}}\right)c^{2}dt'^{2}-2\omega r'^{2}d\phi'dt'-dr'^{2}-r'^{2}d\phi'^{2}-dz'^{2}.\label{eq: Langevin metric}
\end{equation}
The transformation (\ref{eq: trasformazione Langevin}) is both simple
to grasp and highly illustrative of the general covariance of the
GTR as it shows that one can work first in a \textquotedbl{}simpler\textquotedbl{}
frame and then transforming to a more \textquotedbl{}complex\textquotedbl{}
one \cite{key-16,key-17}. As we consider light propagating in the
radial direction ($d\phi'=dz'=0$), the line element (\ref{eq: Langevin metric})
reduces to \cite{key-16}

\begin{equation}
ds^{2}=\left(1-\frac{r'^{2}\omega^{2}}{c^{2}}\right)c^{2}dt'^{2}-dr'^{2}.\label{eq: metrica rotante}
\end{equation}
The EP permits to interpret the line element (\ref{eq: metrica rotante})
in terms of a curved spacetime in presence of a static gravitational
field \cite{key-10,key-15,key-16}. In that way, we obtain a purely
general relativistic interpretation of the pseudo-force experienced
by an observer in a rotating, non-inertial frame of reference \cite{key-16}.
Setting the origin of the rotating frame in the source of the emitting
radiation, we have a first contribution which arises from the ``gravitational
redshift''. It can be directly computed using eq. (25.26) in \cite{key-10}.
In the twentieth printing 1997 of \cite{key-10} that equation is
written as 
\begin{equation}
z\equiv\frac{\Delta\lambda}{\lambda}=\frac{\lambda_{received}-\lambda_{emitted}}{\lambda_{emitted}}=|g_{00}(r'_{1})|^{-\frac{1}{2}}-1.\label{eq: z  MTW}
\end{equation}
Eq. (6) represents the redshift of a photon emitted by an atom at
rest in a gravitational field and received by an observer at rest
at infinity. Here we use a slightly different equation with respect
to eq. (25.26) in \cite{key-10}. In fact, here we are considering
a gravitational field which increases with increasing radial coordinate
$r'$ while eq. (25.26) in \cite{key-10} concerns a gravitational
field which decreases with increasing radial coordinate \cite{key-16}.
Also, we set the zero potential in $r'=0$ instead of at infinity
and we use the proper time instead of the wavelength $\lambda$ \cite{key-16}.
Thus, from eqs. (\ref{eq: metrica rotante}) and (6) we get \cite{key-16}
\begin{equation}
\begin{array}{c}
z_{1}\equiv\frac{\nabla\tau_{10}-\nabla\tau_{11}}{\tau}=1-|g_{00}(r'_{1})|{}^{-\frac{1}{2}}=1-\frac{1}{\sqrt{1-\frac{\left(r'_{1}\right)^{2}\omega^{2}}{c^{2}}}}\\
\\
=1-\frac{1}{\sqrt{1-\frac{v^{2}}{c^{2}}}}\simeq-\frac{1}{2}\frac{v^{2}}{c^{2}}.
\end{array}\label{eq: gravitational redshift}
\end{equation}
In eq. (7) $\nabla\tau_{10}$ is the delay of the emitted radiation,
$\nabla\tau_{11}$ is the delay of the received radiation, $r'_{1}\simeq c\tau$
is the radial distance between the source and the resonant absorber
and $v=r'_{1}\omega$ is the tangential velocity of the resonant absorber
in the rest frame. Hence, we find a first contribution, say $k_{1}=\frac{1}{2}$,
to $k$ \cite{key-16}. We stress again that the power of the EP enabled
us to use a pure general relativistic treatment in the above discussion
\cite{key-16}.

\section{The missing effect: clock synchronization}

Notice that we calculated the variations of proper time $\nabla\tau_{10}$
and $\nabla\tau_{11}$ in the origin of the rotating frame which is
located in the source of the radiation \cite{key-16}. But the detector
of $\gamma$-quanta is moving with respect to the origin in the rotating
frame and with respect to the resonant absorber, see Figure 1. Thus,
the clock in the detector of $\gamma$-quanta must be synchronized
with the clock in the origin. This gives a second contribution to
the redshift and is exactly the point that generated confusion in
our previous work \cite{key-16}. In fact, the authors of \cite{key-17}
claimed that our analysis in \cite{key-16} was incorrect. Let us
clarify this point in an ultimate way. To compute this second contribution
we use eq. (10) of \cite{key-9} which represents the proper time
increment $d\tau$ on the moving clock having radial coordinate $r'$
for values $v\ll c$ 

\begin{equation}
d\tau=dt'\left(1-\frac{r'^{2}\omega^{2}}{c^{2}}\right).\label{eq:secondo contributo}
\end{equation}
Inserting the condition of null geodesics $ds=0$ in eq. (\ref{eq: metrica rotante})
one gets \cite{key-16} 
\begin{equation}
cdt'=\frac{dr'}{\sqrt{1-\frac{r'^{2}\omega^{2}}{c^{2}}}}.\label{eq: tempo 2}
\end{equation}
The positive sign in the square root has been taken because the radiation
is propagating in the positive $r$ direction \cite{key-16}. Combining
eqs. (\ref{eq:secondo contributo}) and (\ref{eq: tempo 2}) one obtains
\cite{key-16}

\begin{equation}
cd\tau=\sqrt{1-\frac{r'^{2}\omega^{2}}{c^{2}}}dr'.\label{eq: secondo contributo finale}
\end{equation}
Eq. (\ref{eq: secondo contributo finale}) is well approximated by
\cite{key-16} 
\begin{equation}
cd\tau\simeq\left(1-\frac{1}{2}\frac{r'^{2}\omega^{2}}{c^{2}}+....\right)dr',\label{eq: well approximated}
\end{equation}
which permits to find the second contribution of order $\frac{v^{2}}{c^{2}}$
to the variation of proper time as \cite{key-16} 
\begin{equation}
c\nabla\tau_{2}=\int_{0}^{r'_{1}}\left(1-\frac{1}{2}\frac{\left(r'_{1}\right)^{2}\omega^{2}}{c^{2}}\right)dr'-r'_{1}=-\frac{1}{6}\frac{\left(r'_{1}\right)^{3}\omega^{2}}{c^{2}}=-\frac{1}{6}r'_{1}\frac{v^{2}}{c^{2}}.\label{eq: delta tau 2}
\end{equation}
$r'_{1}\simeq c\tau$ is the radial distance between the source and
the resonant absorber. Thus, we get the second contribution of order
$\frac{v^{2}}{c^{2}}$ to the redshift as \cite{key-16} 
\begin{equation}
z_{2}\equiv\frac{\nabla\tau_{2}}{\tau}=-k_{2}\frac{v}{c^{2}}^{2}=-\frac{1}{6}\frac{v^{2}}{c^{2}}.\label{eq: z2}
\end{equation}
Then, we obtain $k_{2}=\frac{1}{6}$. Using eqs. (\ref{eq: gravitational redshift})
and (\ref{eq: z2}) the total redshift is \cite{key-16} 
\begin{equation}
\begin{array}{c}
z\equiv z_{1}+z_{2}=\frac{\nabla\tau_{10}-\nabla\tau_{11}+\nabla\tau_{2}}{\tau}=-\left(k_{1}+k_{2}\right)\frac{v^{2}}{c^{2}}\\
\\
=-\left(\frac{1}{2}+\frac{1}{6}\right)\frac{v^{2}}{c^{2}}=-k\frac{v^{2}}{c^{2}}=-\frac{2}{3}\frac{v^{2}}{c^{2}}=0.\bar{6}\frac{v^{2}}{c^{2}}.
\end{array}\label{eq: z totale}
\end{equation}
Eq. (14) is completely consistent with the result $k=0.68\pm0.03$
in \cite{key-2}. Let us clarify the meaning of eq. (\ref{eq: z totale}).
It represents the total energy shift that is detected by the resonant
absorber as it is measured by an observed located in the detector
of $\gamma$-quanta, i.e. located where we have the final output of
the measuring. This is different from the total energy shift that
is detected by the resonant absorber as it is measured by an observed
located in the resonant absorber, which, instead, is given by eq.
(\ref{eq: gravitational redshift}). In fact, the two quantities should
be equal \emph{only }if the detector of $\gamma$-quanta should be
rotating together with the resonant absorber. But the detector of
$\gamma$-quanta is fixed instead. The actual detector (i.e., the
receiver of electromagnetic radiation) is the resonant absorber, whose
resonant line is shifted with respect to the resonant line of the
source. This induces the variation of intensity of resonant gamma-quanta,
passing across this absorber \cite{key-2}. This intensity is measured
by the detector of gamma-quanta, resting outside the rotor system
\cite{key-2}. The latter detector is rather a technical instrument.
It allows experimentalist to judge about the shift of the lines of
the source and the absorber via the measurement of resonant absorption
\cite{key-2}. But the key point is that the shift of the lines of
the source and the absorber that is observed by an observer located
in the rotating resonant absorber \emph{is different} from the shift
of the lines of the source and the absorber that is observed by an
observer located in the fixed detector of $\gamma$-quanta. That difference
is given by the additional factor $-\frac{1}{6}$ in eq. (\ref{eq: z2}),
which comes from clock synchronization. In other words, its theoretical
absence in the works \cite{key-1}-\cite{key-8}, \cite{key-11}-\cite{key-13}
reflected the incorrect comparison of clock rates between a clock
at the origin and one at the of $\gamma$-quanta where we have the
final output of the measuring. This generated wrong claims of invalidity
of relativity theory and/or some attempts to explain the experimental
results through ``exotic'' effects \cite{key-1,key-2,key-11,key-12,key-13,key-17}
which, instead, must be rejected. Let us consider the criticism in
\cite{key-17} about our previous work \cite{key-16}. It verbatim
claims that, ``as the extra energy shift due to the clock synchronization
is of order $10^{-13}$ it cannot be detected by the detectors of
$\gamma$-quanta which are completely insensitive to such a very low
order of energy shifts''. We stress that we are still measuring the
total energy shift by using the resonant absorber instead of by using
the detector of $\gamma$-quanta as it has been claimed in \cite{key-17}.
But the key point is that such a total energy shift measured by an
observer located in the fixed detector of $\gamma$-quanta is different
from the one measured by an observer located in the rotating resonant
absorber. Thus, we have shown that the correct physical interpretation
of a real Mössbauer rotor experiment really represent a new, independent
proof of the GTR contrary to the claims in \cite{key-17}. We also
highlight that the appropriate reference \cite{key-9} has been evoked
for a discussion of the Langevin metric. This is dedicated to the
use of the GTR in Global Positioning Systems (GPS), which leads to
the following interesting realization \cite{key-16}. The correction
of $-\frac{1}{6}$ in eq. (\ref{eq: z2}) is analogous to the correction
that one must consider in GPS when accounting for the difference between
the time measured in a frame co-rotating with the Earth geoid and
the time measured in a non-rotating (locally inertial) Earth centered
frame (and also the difference between the proper time of an observer
at the surface of the Earth and at infinity). Indeed, if one simply
considers the gravitational redshift due to the Earth's gravitational
field, but neglects the effect of the Earth's rotation, GPS would
not work \cite{key-16}! The key point is that the proper time elapsing
on the orbiting GPS clocks cannot be simply used to transfer time
from one transmission event to another. Path-dependent effects must
be indeed taken into due account, exactly like in the above discussion
of clock synchronization \cite{key-16}. In other words, the obtained
correction $-\frac{1}{6}$ in eq. (\ref{eq: z2}) is not an obscure
mathematical or physical detail. It is instead a fundamental ingredient
that must be taken into due account \cite{key-16}. Further details
on the analogy between the results of this paper and the use of the
GTR in Global Positioning Systems have been highlight in \cite{key-16}.

\section{Correct meaning of the WEP and non-viability of the YARK gravitational
theory}

In \cite{key-17} it is claimed that, differently from the GTR, the
so-called YARK (Yarman-Arik-Kolmetskii Gravitational Approach) gravitational
theory \cite{key-18}-\cite{key-24} can explain the result of $k\simeq\frac{2}{3}$
for the total energy shift of the Mössbauer rotor experiment. Actually,
in the above discussion we have shown that, differently from the claims
in \cite{key-17}, a correct physical interpretation of the GTR can
explain the value of $k\simeq\frac{2}{3}$. On the other hand, in
this Section we show that the YARK gravitational theory is in macroscopic
contrast with the WEP. Thus, it results completely non-viable. 

Let us start to observe that the authors of \cite{key-17} claim that,
on one hand, the YARK theory is fully compatible with the WEP. On
the other hand they verbatim claim that ``The real space-time in
a gravitation field remains flat and instead of the geodesic postulate
of GTR, the laws of energy and angular momentum conservation in Minkowskian
space-time are regarded as fundamental''. These two claims are in
macroscopic contrast. In fact, Weinberg \cite{key-25} rigorously
showed that the geodesic motion is NOT a postulate of the GTR, but
a rigorous consequence of the WEP. Before writing the derivation of
this fundamental issue we stress its important consequence. In the
absence of space-time curvature geodesic motion is given by straight
lines \cite{key-26}! But instead, of course, all astrophysical observations
show that the gravitational motion is not given by straight lines
\cite{key-26}. Hence, the only possibility is that space-time is
curved \cite{key-26}. In other words, the YARK's assumption of the
absence of space-time curvature should therefore indicate a macroscopic
violation of the equivalence between the inertial mass and the gravitational
mass \cite{key-26}. But such an equivalence, is instead tested with
the enormous precision of 1 part in $10^{14}$ \cite{key-27,key-28}.
Clearly, considering also the experiments \cite{key-29,key-30,key-31}
etc., it is obvious that YARK's claim of the absence of space-time
curvature is in very strong contrast with tons of data collected in
more than a century. Now, let us show that the WEP implies that test
masses must follow geodesic lines. Notice that in the following derivation
we closely follow \cite{key-25,key-26}. Let us start supposing that
no particles are accelerating in the neighborhood of a point-event
with respect to a freely falling coordinate system $\left(X^{\mu}\right)$
\cite{key-25,key-26}. Putting $T=X^{0}$ one writes down the following
equation that is locally applicable in free fall \cite{key-25,key-26}

\begin{equation}
\frac{d^{2}X^{\mu}}{dT^{2}}=0.\label{eq: free fall}
\end{equation}
Using the chain rule one gets \cite{key-25,key-26}

\begin{equation}
\frac{dX^{\mu}}{dT}=\frac{dx^{\nu}}{dT}\frac{\partial X^{\mu}}{\partial x^{\nu}}.\label{eq: chain rule}
\end{equation}
Differentiating eq. (\ref{eq: chain rule}) with respect to $T$ one
gets \cite{key-25,key-26} 
\begin{equation}
\frac{d^{2}X^{\mu}}{dT^{2}}=\frac{d^{2}x^{\nu}}{dT^{2}}\frac{\partial X^{\mu}}{\partial x^{\nu}}+\frac{dx^{\nu}}{dT}\frac{dx^{\alpha}}{dT}\frac{\partial^{2}X^{\mu}}{\partial x^{\nu}\partial x^{\alpha}}.\label{eq: Differentiating}
\end{equation}
Combining eqs. (\ref{eq: free fall}) and (\ref{eq: Differentiating})
one immediately gets \cite{key-25,key-26} 
\begin{equation}
\frac{d^{2}x^{\nu}}{dT^{2}}\frac{\partial X^{\mu}}{\partial x^{\nu}}=-\frac{dx^{\nu}}{dT}\frac{dx^{\alpha}}{dT}\frac{\partial^{2}X^{\mu}}{\partial x^{\nu}\partial x^{\alpha}}.\label{eq: nullo}
\end{equation}
Multiplying both sides of eq. (\ref{eq: nullo}) by $\frac{\partial x^{\lambda}}{\partial X^{\mu}}$
one gets \cite{key-25,key-26}

\begin{equation}
\frac{d^{2}x^{\lambda}}{dT^{2}}=-\frac{dx^{\nu}}{dT}\frac{dx^{\alpha}}{dT}\left[\frac{\partial^{2}X^{\mu}}{\partial x^{\nu}\partial x^{\alpha}}\frac{\partial x^{\lambda}}{\partial X^{\mu}}\right].\label{eq: Multiplying}
\end{equation}
Setting $t=x^{0}$ and using again the chain rule, $T$ can be eliminated
in favor of the coordinate time $t$ \cite{key-25,key-26} 
\begin{equation}
\frac{d^{2}x^{\lambda}}{dt^{2}}=-\frac{dx^{\nu}}{dt}\frac{dx^{\alpha}}{dt}\left[\frac{\partial^{2}X^{\mu}}{\partial x^{\nu}\partial x^{\alpha}}\frac{\partial x^{\lambda}}{\partial X^{\mu}}\right]+\frac{dx^{\nu}}{dt}\frac{dx^{\alpha}}{dt}\frac{dx^{\lambda}}{dt}\left[\frac{\partial^{2}X^{\mu}}{\partial x^{\nu}\partial x^{\alpha}}\frac{\partial x^{0}}{\partial X^{\mu}}\right].\label{eq: quasi geodetiche}
\end{equation}
Recalling that the bracketed terms involving the relationship between
local coordinates $X$ and general coordinates $x$ are functions
of the general coordinates, eq. (\ref{eq: quasi geodetiche}) gives
immediately the geodesic equation of motion using the coordinate time
$t$ as parameter \cite{key-25,key-26} 
\begin{equation}
\frac{d^{2}x^{\lambda}}{dt^{2}}=-\Gamma_{\nu\alpha}^{\lambda}\frac{dx^{\nu}}{dt}\frac{dx^{\alpha}}{dt}+\Gamma_{\nu\alpha}^{0}\frac{dx^{\nu}}{dt}\frac{dx^{\alpha}}{dt}\frac{dx^{\lambda}}{dt},\label{eq: geodetiche rispetto a t}
\end{equation}
which is equivalent to the standard geodesic equation written in terms
of the scalar parameter $s$ \cite{key-25,key-26} 
\begin{equation}
\frac{d^{2}x^{\lambda}}{ds^{2}}=-\Gamma_{\nu\alpha}^{\lambda}\frac{dx^{\nu}}{ds}\frac{dx^{\alpha}}{ds}.\label{eq: geodetiche rispetto ad s}
\end{equation}
Clearly, based on the extreme precision on which the WEP is today
tested and verified, the demonstration that we have reviewed here
- i.e. that geodesic motions arise from the WEP - ultimately rules
out the YARK theory. Infact, that theory is founded on the absence
of curvature \cite{key-17}-\cite{key-24}. 

In addition, the authors of \cite{key-17} claim that YARK theory
permits to localize the gravitational energy. In their opinion, it
should remain a non-vanishing quantity in all plausible frames of
reference. This should permit to write down, explicitly, a stress-energy
tensor for the gravitational field. These claims are again in macroscopic
contrast with the WEP \cite{key-10}. Another consequence of the WEP
is indeed that we can always find in any given locality a reference's
frame (the local Lorentz reference's frame) in which ALL local gravitational
fields are null \cite{key-10}. No local gravitational fields means
no local gravitational energy-momentum and, in turn, no stress-energy
tensor for the gravitational field \cite{key-10}. Thus, these are
other strong reasons for which the YARK theory of gravity is non-viable
and must be ultimately rejected.

\section{Conclusion remarks}

In \cite{key-16} we used the EP, which states the equivalence between
the gravitational \textquotedbl{}force\textquotedbl{} and the \emph{pseudo-force}
experienced by an observer in a non-inertial frame of reference (included
a rotating frame of reference), to reanalyze the theoretical framework
of the Mössbauer rotor experiment directly in the rotating frame of
reference. We used a\emph{ }general relativistic treatment. We have
shown that previous analyses missed an important effect of clock synchronization
and that the correct general relativistic prevision in the rotating
frame gives a pre-factor $k\simeq\frac{2}{3}$ for the total energy
shift of the Mössbauer rotor experiment \cite{key-16}. This is in
perfect agreement with new experimental results \cite{key-1,key-2}.
The effect of clock synchronization has been indeed missed in various
papers in the literature \cite{key-1}-\cite{key-8}, \cite{key-11}-\cite{key-13},
with some subsequent claim of invalidity of the relativity theory
and/or some attempts to explain the experimental results through ``exotic''
effects \cite{key-1,key-2,key-11,key-12,key-13,key-17}. The general
relativistic interpretation in \cite{key-16} showed, instead, that
the new experimental results of the Mössbauer rotor experiment are
a new, strong and independent, proof of the GTR. 

In the recent work \cite{key-17}, it is claimed that the general
relativistic treatment in \cite{key-16} , with the additional effect
due to clock synchronization, cannot explain the extra energy shift
in the Mössbauer rotor experiment. The extra energy shift due to the
clock synchronization is indeed of order $10^{-13}$ and cannot be
detected by the detectors of $\gamma$-quanta which are completely
insensitive to such a very low order of energy shifts \cite{key-17}.
In addition, in \cite{key-17}, it is also claimed that the extra
energy shift can be explained in the framework of the so-called YARK
gravitational theory \cite{key-17}-\cite{key-24}. In the opinion
of the authors of \cite{key-17}-\cite{key-24} this new theory should
replace the GTR as the correct theory of gravity.

In this paper we have shown that the theoretical analysis in \cite{key-16}
has been misunderstood by the authors of \cite{key-17}. In fact,
in \cite{key-16} it has been shown that electromagnetic radiation
launched by the central source of the apparatus is redshifted of a
quantity $0.\bar{6}\frac{v^{2}}{c^{2}}$ when arriving to the detector
of $\gamma$-quanta. This holds independently by the issue that the
original photons are detected by the resonant absorber which, in turns,
triggers the $\gamma$-quanta which arrive to the final detector.
In other words, the result of \cite{key-16} is purely theoretical
and is completely independent of the way the experiment is concretely
realized. In the present work we have shown that, with some clarification,
the results in \cite{key-16} hold also when we consider the various
steps of the concrete detection. In that case, the resonant absorber
detects the energy shift and the separated detector of $\gamma$-quanta
merely measures the resulting intensity. In addition, in this paper
we have also shown that the YARK gravitational theory is in macroscopic
contrast with geodesic motion and, in turn, with the WEP. This is
in contrast with another claim of the authors of \cite{key-17} which
states that the YARK gravitational theory arises from the WEP. Therefore,
the YARK gravitational theory has to be ultimately rejected. We have
also corrected the confusion of the authors of \cite{key-17} concerning
their claims about the possibility to localize the gravitational energy
and, in turn, to define a stress-energy tensor for the gravitational
field. In fact, we show that these claims are still in macroscopic
contrast with the WEP.

Again we stress that we dedicate the results in this paper to the
100th anniversary of Albert Einstein's presentation of the complete
GTR to the Prussian Academy.

\section{Acknowledgements }

The author thanks the anonymous Editor and the anonymous Referee for
appreciating this work and for useful comments.

\end{document}